# Coherence analysis of beamline and simulation of coherent X-ray diffraction imaging based on a coherent modes model

Han Xu,[a] Zhongzhu Zhu,[a] Peng Liu,[a] Yuhui Dong[a] and Liang Zhou[a*]

[a]*Beijing Synchrotron Radiation Facility, Institute of High Energy Physics, Chinese Academy of Sciences, Beijing, People's Republic of China.*

[*]zhouliang@ihep.ac.cn.

# Abstract

The evolution from 3$^{rd}$ to 4$^{th}$ generation synchrotron radiation (SR) sources significantly enhanced their coherence, making coherent scattering techniques such as coherent X-ray diffraction imaging (CXDI) and X-ray photon correlation spectroscopy (XPCS) more accessible. We take advantage of a coherent modes decomposition and propagation model to carry out coherence analysis and simulations of CXDI experiments for the coherent Hard X-ray Coherent Scattering (HXCS) beamline at High Energy Photon Source.

# 1. Introduction

Synchrotron radiation (SR) has advanced from 3rd to 4th generation sources with the development of multi-bend achromat lattices (Hettel, 2014), which allow electron emittance to be reduced to enhance the brilliance and coherence of the SR source. For example, the electron emittance for the High Energy Photon Source (HEPS) is about 34 pm·rad (Jiao et al., 2018). Enhanced coherence at 4th generation SR sources benefit coherent scattering techniques such as coherent X-ray diffraction imaging (CXDI) (Miao et al., 2015; Pfeiffer et al., 2006) and X-ray photon correlation spectroscopy (XPCS) (Madsen et al., 2015). The quality of CXDI and the signal-to-noise ratio of XPCS strongly depend on the coherence of the X-ray beams at the sample, and it is therefore necessary to analyze the coherence modulation of X-ray beamlines. Furthermore, coherent scattering experiments should be simulated by wave optics to explore the performance of 4th generation SR beamlines.

Recently, simulations of SR beamline coherent scattering experiments have utilized wave optics methods. For example, the Monte Carlo multi-electron wavefronts propagation method considers the emittance and propagation of X-ray wavefronts for every sampled electron, and the coherence effect is calculated statistically (Chubar et al., 1998; Chubar et al., 2013). Based on this method, simulations of coherent scattering experiments, such as coherent X-ray scattering and XPCS, have been performed on 3rd generation SR sources (NSLS-II and BNL, Chubar et al., 2017 and Chubar et al., 2020). Although accuracy could be ensured, the demands on computational resources reduced the flexibility of this method, leading to exploration of other models (Zhu et al., 2021). Taking advantage of the enhanced coherence of 4th generation SR sources, recently-developed coherent modes decomposition and propagation methods have become more efficient. Furthermore, accurate coherent fraction of X-ray beam can be provided by this method.

Based on statistical optics, the coherence of an X-ray beam can be fully described by the cross-spectral density (CSD). The expansion of CSD results in orthonormal coherent modes (Ostrovsky *et al*., 2006). Taking advantage of the enhanced coherence of 4$^{th}$ generation SR, the number of required coherent modes for coherence analysis is ~$10^2$ (Glass *et al.*, 2017), which is much more efficient than for wavefronts (~$10^5$ for the Monte Carlo multi-electron method). Several tools have been developed for coherent mode analysis. For example, the CSD of the ESRF-EBS (European Synchrotron Radiation Facility-Extremely Brilliant Source) was calculated using the brightness convolution theorem prior to coherent modes decomposition (Glass *et al.*, 2017; Sanchez del Rio *et al.*, 2019). X-ray tracing software is available that incorporates a coherent mode calculation tool based on wavefronts (Khubbutdinov *et al.*, 2019).

In this study, coherence analysis and simulations of CXDI have been conducted for the Hard X-ray Coherent Scattering (HXCS) beamline at an HEPS using a coherent modes decomposition and propagation model. Firstly, a multi-layer singular-value decomposition (SVD) method combined the accuracy of the Monte Carlo multi-electron method and the high efficiency of the coherent modes method. Then, taking advantage of the high efficiency of coherent mode propagation, X-ray beams of differing partial coherence were constructed for CXDI simulations to help define the optical design requirements for CXDI experiments at HXCS beamlines.

## 2. Model description

### 2.1 Coherent modes decomposition of SR X-rays

Usually, coherent modes are decomposed from the CSD $W(x_1, x_2; y_1, y_2, \omega)$ by eigenvalue algorithms such as the Arnoldi method (Ostrovsky *et al.*, 2006). And the CSD is calculated from the wavefronts of the radiated electric fields:

$$W(x_1, x_2; y_1, y_2, \omega) = \langle E^*(x_1, y_1; \omega) E(x_2, y_2; \omega) \rangle, \tag{1}$$

$$W(x_1, x_2; y_1, y_2, \omega) = \sum_m \rho_m \varphi_m^*(x_1, y_1; \omega) \varphi_m(x_2, y_2; \omega), \tag{2}$$

where $E$ represents the wavefronts radiated by the electron beam (Glass *et al.*, 2017), $\varphi_m(r)$ is the coherent mode, and $\rho_m$ is the eigenvalue.

The SVD method provides an alternative strategy for coherent modes decomposition without calculating the CSD:

$$A = U \Sigma V, \tag{3}$$

where $A$ is a data matrix with $Nx \times Ny$ rows and $n$ columns. Each column contains a wavefront ($Nx \times Ny$ pixels, reshaped to one dimension), $A = [E_0, E_1, …, E_n]$. $U$ is $AA^+$ ($A^+$ is the Hermitian transpose of $A$) which is equal to the CSD (($Nx \times Ny)^2$ pixels). $V$ is $A^+A$ ($n^2$ pixels). $\Sigma$ is a diagonal matrix of eigenvalues. The SVD method decomposes the matrix $V$ instead of the much larger matrix $U$ (CSD). If the number of wavefronts $n$ is smaller than $Nx \times Ny$, the SVD method is more efficient than CSD decomposition. However, for the Monte Carlo multi-electron method, the number of electrons is $\sim 10^5$ and the spatial dimension of the wavefront is $\sim 10^4$–$10^6$ (Sanchez del Rio *et al.*, 2019), in which case the SVD method does not offer an advantage.

Here, a multi-layer SVD method was developed. Firstly, $A$ is divided into $[A_1, A_2, …, A_k]$ along the direction of column, and the dimension of $A_i$ is $(Nx \times Ny) \times (n/k)$. Secondly, SVD was performed to generate coherent modes for each part of $[A_0, A_1, …, A_k]$:

$$A_i = \left[ E_{i1}, E_{i2}, …, E_{i\left(\frac{n}{k}\right)} \right] = \left[ \sqrt{\rho_{i1}} \varphi_{i1}, \sqrt{\rho_{i2}} \varphi_{i2}, …, \sqrt{\rho_{it}} \varphi_{it} \right], \tag{4}$$

where $t$ is the truncation number of the coherent modes decomposition. By using the decomposed coherent modes instead of wavefronts to construct $A$, and dimension of column changes from $n$ to $t \times k$. Then, SVD can be performed.

SVD can be performed individually for each part of $[A_0, A_1, …, A_k]$. Thus, the structure of this method is conducive to parallel computing. More importantly, this method takes full advantage of the enhanced coherence of 4$^{th}$ generation SR sources since high-coherence sources strongly reduce $t$ and therefore $t \times k$. The validity of this method has been verified by comparison with Synchrotron Radiation Workshop (SRW) and Comsyl software (see **S1. Verification of coherent modes decomposition** and **Figure. S1-S4**).

## 2.2 Propagation of coherent modes

Different propagators were applied for various optical layouts.

The Fresnel diffraction integral was usually used to propagate the coherent modes from one optic ($x_1$, $y_1$) to another ($x_2$, $y_2$). For one coherent mode $\varphi(x, y)$,

$$\varphi(x_2, y_2) = \frac{k e^{ikd}}{2\pi i d} \iint_{-\infty}^{+\infty} \varphi(x_1, y_1) e^{\frac{ik}{2d}[(x_2-x_1)^2 + (y_2-y_1)^2]} dx_1 dy_1, \qquad (5)$$

where $k$ is $\frac{2\pi}{\lambda}$ ($\lambda$ is wavelength), $d$ is the distance between two optical elements (Voelz, 2011). The Fresnel diffraction integral was calculated by a double fast Fourier transform (*D-FFT*) algorithm:

$$\varphi(x_2, y_2) = IFFT\{FFT[\varphi(x_1, y_1)] \times FT[h(x_2, y_2)]\}, \qquad (6)$$

where *FT* is Fourier transform, *IFFT* is inverse fast Fourier transform, and

$$FT[h(x_1, y_1)] = FT\left[\frac{k e^{ikd}}{2\pi i d} e^{\frac{ik(x_2^2 + y_2^2)}{2d}}\right] = e^{ikd\left[1 - \frac{2\pi^2}{k^2}(q_x^2 + q_y^2)\right]}. \qquad (7)$$

For optics like advanced-KB mirror (AKB), the distance between mirrors could be quite small (see **S3. Propagation of coherent modes based on the optical layout of 100 nm** **focusing** and **Figure. S7**). The coherent modes were propagated by angular spectrum method (Shen *et al.*, 2006). For one coherent mode $\varphi(x, y)$,

$$\varphi(x_2, y_2) = \iint_{-\infty}^{+\infty} \varphi(q_x, q_y) e^{id\sqrt{k^2 - 4\pi^2(q_x^2 + q_y^2)}} e^{i2\pi(x_2 q_x + y_2 q_y)} dq_x dq_y, \qquad (8)$$

where,

$$\varphi(q_x, q_y) = \iint_{-\infty}^{+\infty} \varphi(x_1, y_1) e^{-i2\pi(x_1 q_x + y_1 q_y)} dx_1 dy_1, \qquad (9)$$

is the Fourier transform of $\varphi(x_1, y_1)$. The angular spectrum was calculated by the *FFT-AS* method:

$$\varphi(x_2, y_2) = IFFT\{FFT[\varphi(x_1, y_1)] \times e^{id\sqrt{k^2 - 4\pi^2(q_x^2 + q_y^2)}}\}. \qquad (10)$$

For sub-micrometer focusing, the small focal length and requirement for high axial resolution could greatly increase the spatial dimension of the optical plane and thus computation time. For the Fresnel diffraction integral, the single-*FFT* algorithm is,

$$\varphi(x_2, y_2) = F_0 \times FT[\varphi(x_1, y_1) \times F], \qquad (11)$$

where $F_0$ is $\frac{ke^{ikd}}{2\pi id} e^{\frac{ik}{2d}(x_2^2 + y_2^2)}$, $F$ is $e^{\frac{ik}{2d}(x_1^2 + y_1^2)}$. This problem has been solved (Bluestein, 1970; Rabiner *et al.*, 1969) by calculating this *FT* with the chirp-z transform (CZT) method. Its application to wave optics has recently been reported (Hu *et al.*, 2020). Here, this method is applied for sub-micrometer focusing.

The correctness of these propagators was verified by comparison with SRW (Chubar *et al.*, 1998; see **S2. Verification of propagators** and **Figure. S5 and S6**). For the optical layout described in this article, the *D-FFT* Fresnel diffraction integral was used for the propagation of coherent modes. Another optical layout of HXCS beamlines (100 nm focusing) has been provided as an example of the performance of other propagators (see **S3. Propagation of coherent modes based on the optical layout of 100 nm focusing** and **Figure. S7**).

**2.3. The CXDI model**

The simulation of CXDI was based on the coherent modes and the sample. Using the projection approximation, a sample can be described by a transmission function $T(x, y)$ (Chubar *et al.*, 2017). When a coherent mode $\varphi_i(x, y)$ illuminates the sample, the exit coherent mode can be represented by $T(x, y)\varphi_i(x, y)$, which then propagates to the detector. For traditional plane wave CXDI, the distance $d$ between the sample ($x_1$, $y_1$) and the detector ($x_2$, $y_2$) should be large enough to allow the propagation to be treated as Fraunhofer diffraction:

$$\varphi(x_2, y_2) = \frac{ke^{ikd}}{2\pi i d} e^{\frac{ik(x_2^2 + y_2^2)}{2d}} \iint_{-\infty}^{+\infty} T(x_1, y_1)\varphi(x_1, y_1) e^{\frac{-ik(x_1 x_2 + y_1 y_2)}{d}} dx_1 dy_1. \quad (12)$$

Thus, $\varphi(x_2, y_2)$ is the Fourier transform of $\varphi(x_1, y_1)$. Here, the *FFT* was used:

$$\varphi(q_x, q_y) = FFT[T(x, y)\varphi(x, y)]. \quad (13)$$

Therefore, the diffraction pattern at the detector can be calculated by summing the intensity of all the propagated coherent modes:

$$I(q_x, q_y) = \sum_m \rho_m |\varphi_m(q_x, q_y)|^2 = \sum_m \rho_m |FFT[T(x_1, y_1)\varphi_m(x_1, y_1)]|^2 \quad (14)$$

where $\varphi_m(r)$ is the coherent mode and $\rho_m$ is the eigenvalue. Finally, by assuming plane waves, $T(x, y)$ can be reconstructed by phase retrieval algorithms.

## 3. Simulations

### 3.1. SR source

For the HEPS, the horizontal electron emittance was greatly reduced (34 pm·rad) and the energy spread was $1.061 \times 10^{-3}$. Based on the parameters of the electron beam and undulator (**Table 1**), coherent modes decomposition of the HXCS beamline was performed to analyze the coherence. The SR source was simulated by $8 \times 10^4$ electrons using the Monte Carlo method, and the emitted wavefronts were calculated using SRW and a distance of 20 m. Depending on the wavefronts, multi-layer SVD of coherent modes was performed. The results for coherent modes with $m = 0\sim5$ are shown in **Figure 1**.

The following describes the results of multi-layer SVD. For the HXCS beamline SR source, $n = 8\times10^4$, $k = 20$, $n/k = 4000$, $t = 200$, and $t\times k = 4\times10^3$. Firstly, 20 processes (in 20 cores) were used for parallel computation of the SVD of $A_i$ (i = 1~20). Secondly, $A$ was reconstructed using decomposed coherent modes, and the dimension of the column changed from $n = 8\times10^4$ to $t\times k = 4\times10^3$. Thirdly, the SVD of $A$ was performed at the root process to generate the final coherent modes and eigenvalues. The coherent fraction (*CF*) can be defined by the occupation of the first coherent mode (Khubbutdinov *et al.*, 2019). As shown in Figure 1, the calculated *CF* of the SR source for the HXCS beamline was 20.7%, and the occupation of coherent modes with $m = 0$ to $m = 4$ already exceeded 50%, which demonstrated the high coherence of the HEPS.

**3.2. Coherence analysis of the beamline**

An optical layout of HXCS designed for XPCS and traditional plane wave CXDI is shown in **Figure 2**. The sample was placed at 2.7 m from the focus and the coherence and coherent flux were adjusted by the secondary source slit. The secondary source was focused by a 4:3 focus system (object distance = 40 m, imaging distance = 29.3 m). Based on the data in **Table 1**, the SR photon source divergence was calculated to be 4.3 $\mu$rad (horizontal direction) and 3.8 $\mu$rad (vertical direction). Thus, the size of the propagated X-ray beam incident on the lens was 172 $\mu$m (4.3 $\mu$rad×40 m). The diameter of the lens was set to 600 $\mu$m (>3$\sigma$) to receive most of the incident X-ray beam, and its focal length was 16.91 m.

The X-ray beam at the sample was simulated by propagating the coherent modes ($m = 0\sim29$) from the source to the sample. Firstly, the secondary source was focused by the ideal lens.

For the ideal lens, the phase transformation was $e^{\frac{ik(x^2+y^2)}{2f}}$, where $f$ is the focal length. As shown in **Figure 3**, the intensity, CSD$x$ ($W(x_1, x_2; y_1 = 0, y_2 = 0, \omega)$), and CSD$y$ ($W(x_1 = 0, x_2 = 0; y_1, y_2, \omega)$) of the secondary source were calculated by propagated coherent modes based on **Equation 2**. Based on **Figure 3a**, the size of the secondary source (horizontal×vertical) was 8.81 $\mu$m×4.14 $\mu$m, consistent with the theoretical value (7.69 $\mu$m×4.10 $\mu$m). The coherence lengths (FWHM) were calculated to be 3.87 $\mu$m×6.08 $\mu$m (**Figure 3b,c**). Secondly, the secondary source slit (*SSS*) was used to modify the partial coherence of the X-ray beam. The size of the slit was adjustable from 21.0 $\mu$m×21.0 $\mu$m to 12.6 $\mu$m×12.6 $\mu$m. Finally, the modified coherent modes were propagated from the secondary source to the sample (10 $\mu$m×10 $\mu$m).

### 3.3. Simulation of CXDI

Simulations of plane wave CXDI were carried out with partial coherence represented by the *CF*. Coherent modes modulated by optical elements are usually no longer orthonormal (Sanchez del Rio *et al.*, 2019), and so re-diagonalization of the X-ray beams at the sample was performed to calculate the new coherent modes. SVD was deemed to be an appropriate and highly efficient method here since $n = 30$. The spatial dimension of data matrix *A* for SVD was ($Nx \times Ny$)×$n$ (horizontal×vertical). Column *m* of *A* consisted of $\sqrt{\rho_m}\varphi_m$ (reshaped to one dimension). As described above, SVD decomposes matrix $A^+A$ ($n \times n$). The coherent modes after re-diagonalization are shown in **Figure 4d–f** (*SSS* = 21 $\mu$m, 10 $\mu$m×10 $\mu$m at the sample). SVD showed that the *CF* of the X-ray beams at the sample increased from 89.0% to 95.6% as the width of the *SSS* decreased from 21.0 $\mu$m to 12.6 $\mu$m.

The standard sample used for simulations was a part of the Siemens star (**Figure 5a**, binary sample, created by PyNX, 10 $\mu$m×10 $\mu$m, 256×256 pixels, Favre-Nicolin *et al.*, 2020). The spatial dimension of the diffraction pattern was 800×800 pixels (*FFT* of the sample: **Figure 5b**, 60 $\mu m^{-1}$×60 $\mu m^{-1}$), and the oversampling ratio was 9.8, which satisfies the requirements of CXDI for accurate sample reconstruction. The sample was reconstructed by the Hybrid Input-Output (HIO) algorithm (Fienup, 1982) and Error Reduction (ER) method (Fienup, 1982). The HIO algorithm can avoid the local-minimum stagnation problem, and the ER algorithm guarantees convergence of the reconstruction.

The results of the reconstruction are shown in **Figure 6a–f**. It was observed that the quality of the reconstruction increases with coherence. Although the coherence length covered the size of the sample when the coherent fraction is 89.0% (10.3 $\mu$m×18.0 $\mu$m), an acceptable reconstruction could not be achieved until the coherent fraction increased to 94.7%. To increase the quality of the CXDI reconstruction, an X-ray beam with a higher *CF* was required. However, the flux of the X-ray beam was reduced by decreasing the width of the secondary source slit. Thus, the *CF* and flux of the X-ray beam must be balanced. As shown in **Figure 7a**, the relative flux of the X-ray beam (normalized by the flux from the source) at the sample decreased to 3.82% when the *CF* = 94.7%. Further increasing the *CF* to 95.6% required a smaller slit (12.6 $\mu$m), which reduced the intensity to 3.47%.

## 4. Conclusion

In summary, based on a recently developed coherent modes decomposition and propagation model, coherence analysis and traditional plane wave CXDI simulations have been conducted for a typical HXCS beamline at a 4[th] generation SR HEPS. A multi-layer SVD method was developed to perform effective coherent modes decomposition of highly coherent X-rays from a 4[th] generation SR source. Taking advantage of coherent mode propagation, X-ray beams were simulated with different partial coherence values at the sample. The effect of partial coherence on plane wave CXDI experiments was analyzed.

**Tables and Table captions:**

**Table 1**
The parameters of the electron beam and undulator of the HXCS and NAMI (Hard X-ray Nanoprobe Multimodal Imaging) beamline.

| Beamline | HXCS | NAMI |
|---|---|---|
| Current | 0.2 A | 0.2 A |
| Energy | 6 GeV | 6 GeV |
| Energy spread | $1.061 \times 10^{-3}$ | $1.100 \times 10^{-3}$ |
| Electron beam size, $\sigma_x$ | 9.334 μm | 9.334 μm |
| Electron beam size, $\sigma_y$ | 2.438 μm | 2.438 μm |
| Electron beam divergence, $\sigma_{x'}$ | 3.331 μrad | 3.331 μrad |
| Electron beam divergence, $\sigma_{y'}$ | 1.275 μrad | 1.275 μrad |
| Undulator length | 4 m | 3.932 m |
| Number of undulator periods | 201 | 174 |

**Figures and Figure captions:**

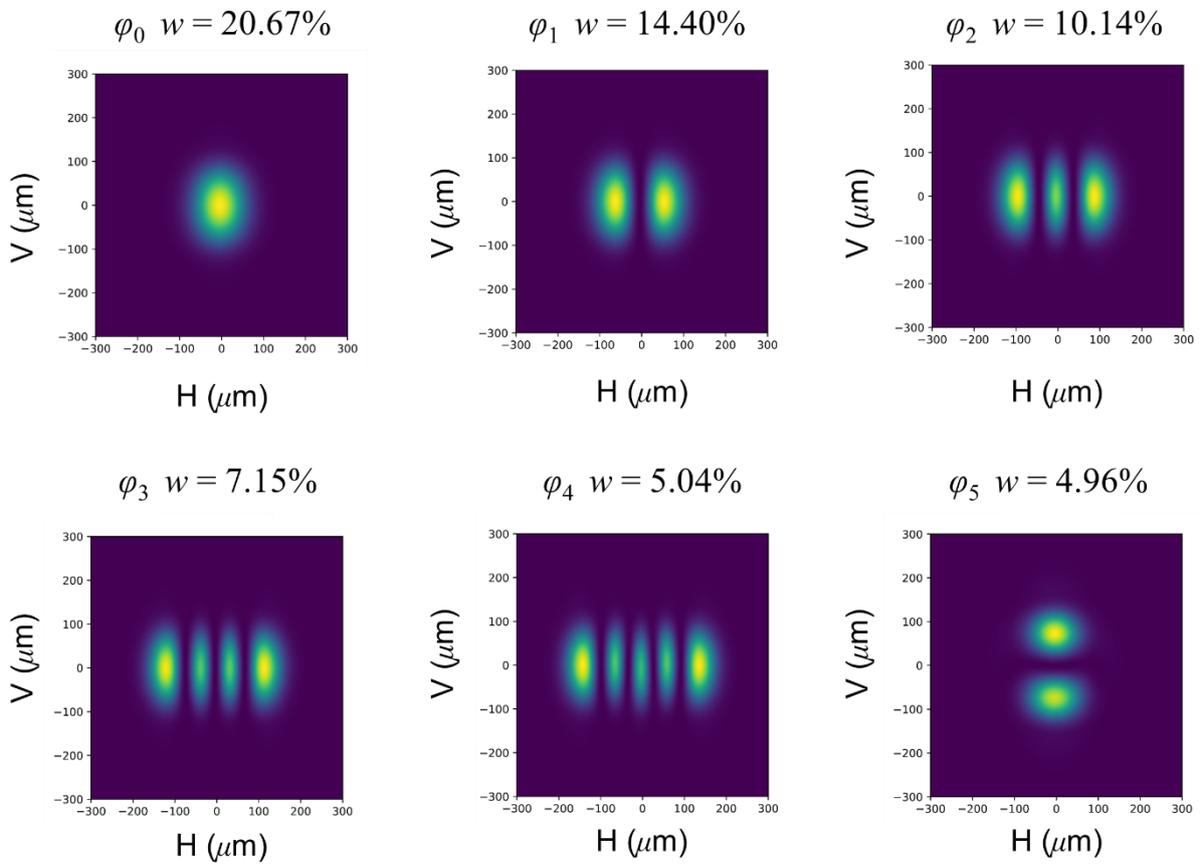

Figure 1. The first six coherent modes ($\varphi_0$~$\varphi_5$) and normalized weights ($w$) from the HEPS.

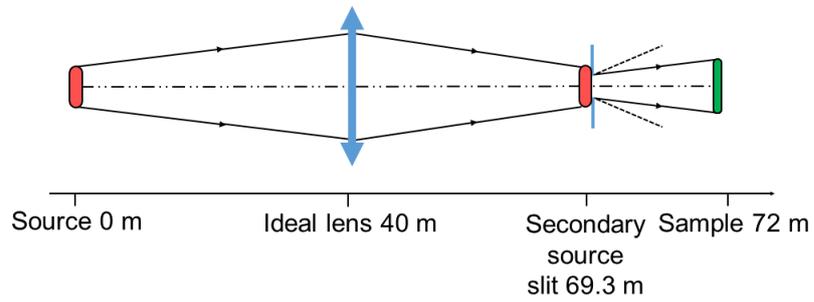

Figure 2. HXCS optical layout designed for XPCS and traditional plane wave CXDI.

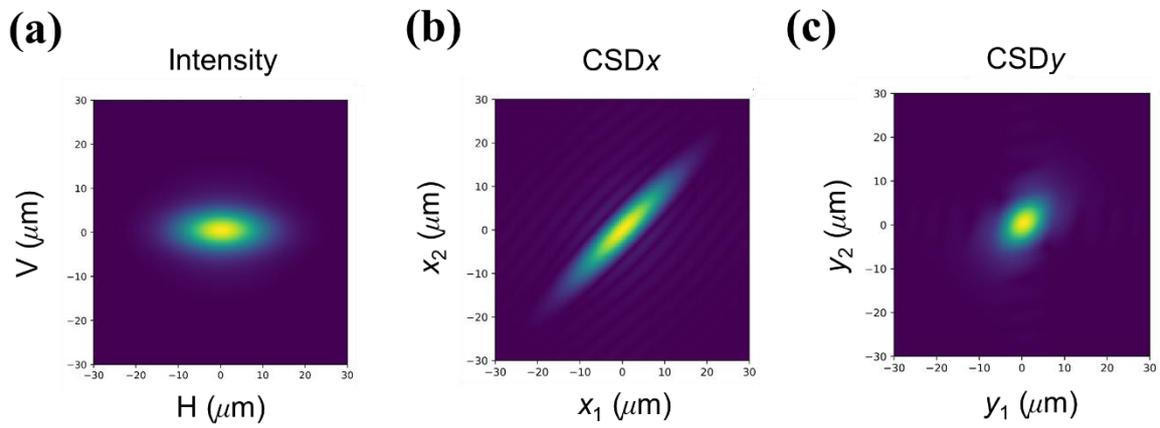

Figure 3. (a) intensity, (b) CSD$x$, and (c) CSD$y$ of the secondary source.

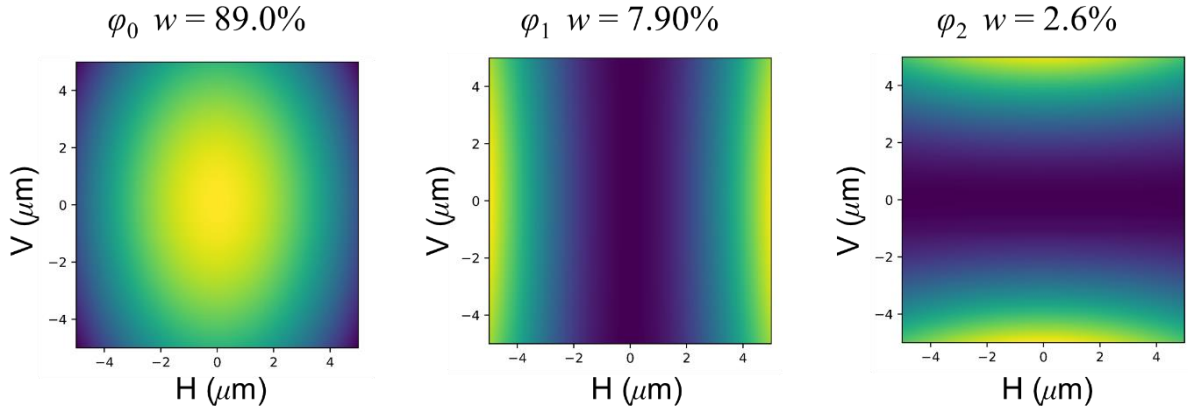

Figure 4. The coherent modes the X-ray beam at the sample (m = 0~2, after re-diagonalization, secondary slit size = 21 $\mu$m).

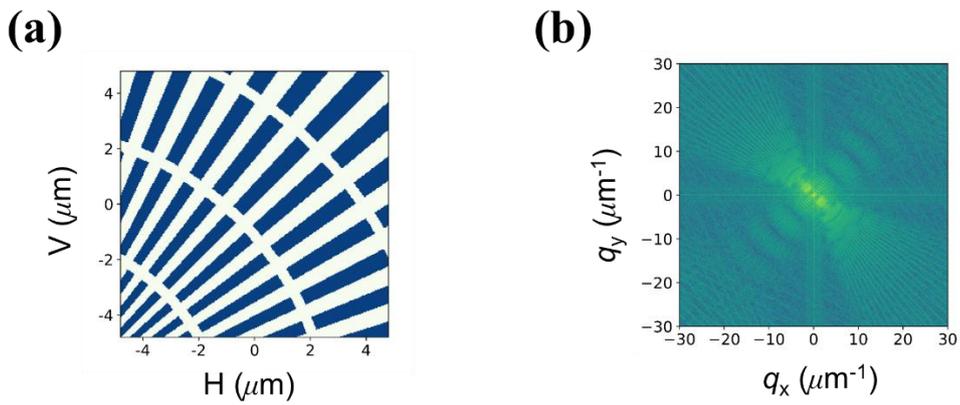

Figure 5. (a) The sample used for *pw*-CXDI simulations. (b) The diffraction pattern of from sample.

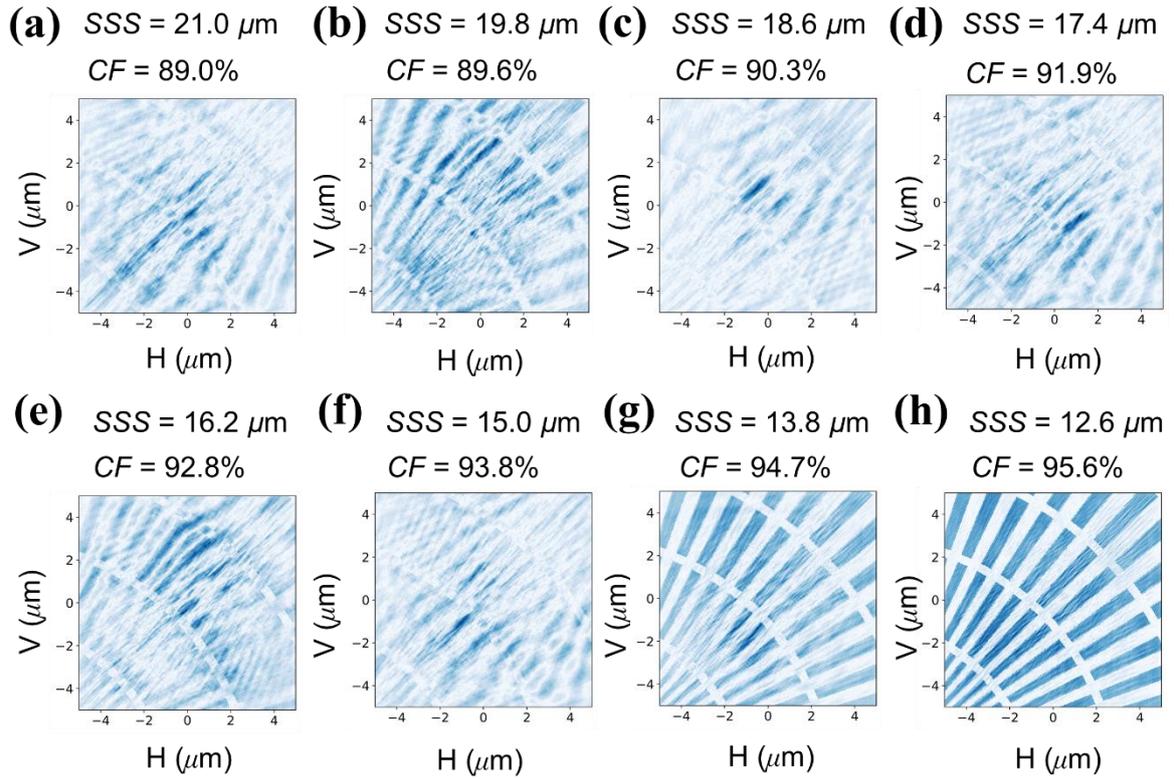

Figure 6. Reconstruction results for CXDI simulations with the coherent fraction *(CF)* ranging from (a) 89.0% to (g) 95.6%.

## S1. Verification of coherent modes decomposition.

The intensity, CSD$x$ and CSD$y$ were calculated and compared by the decomposed coherent modes and SRW software. the distance between the optical plane and source is 10 m. The results of HXCS and NAMI (Hard X-ray Nanoprobe Multimodal Imaging) beamlines were shown here.

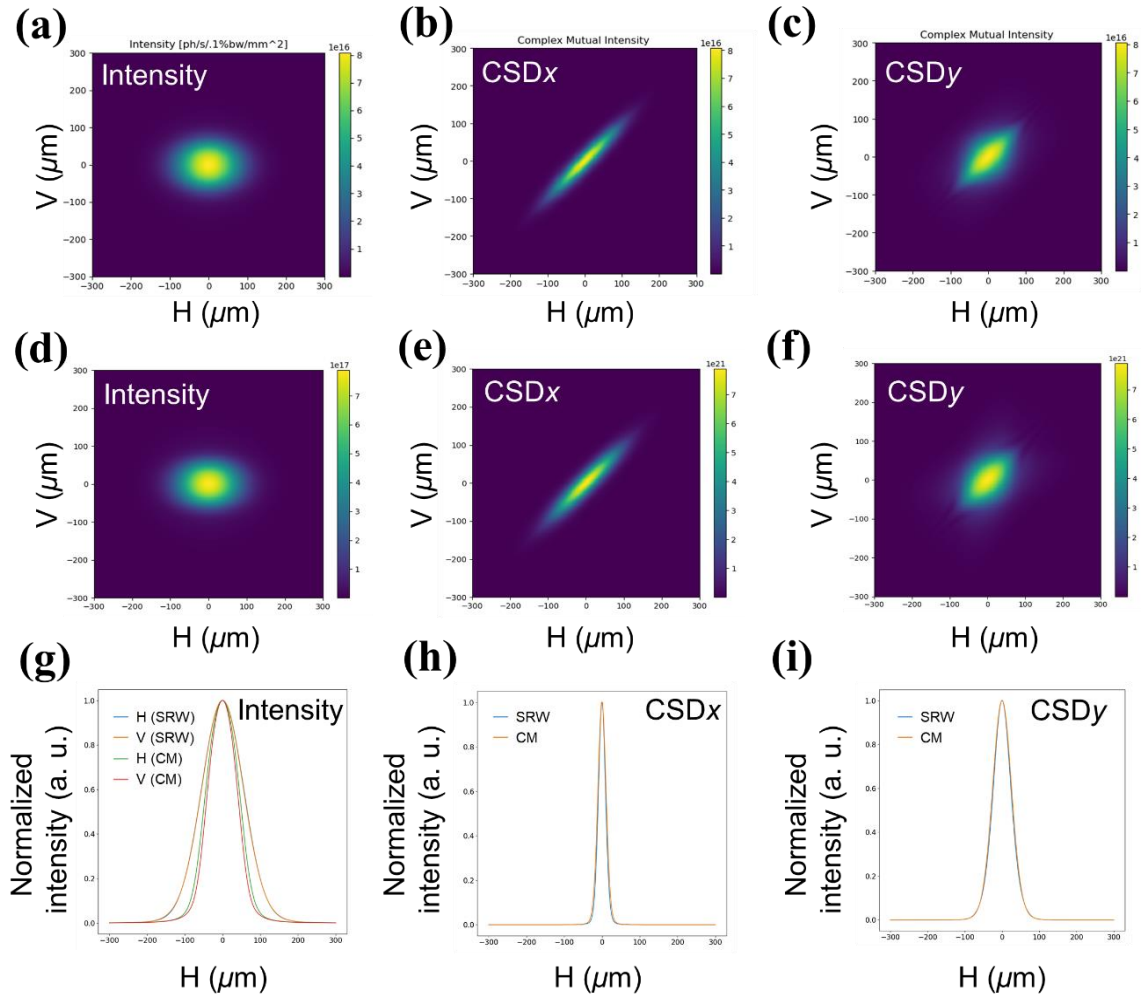

**Figure. S1.** The comparison based on NAMI beamline (10keV, **Table 1**, **column NAMI**). (a) The intensity calculated by SRW. (b) The CSD$x$ calculated by SRW. (c) The CSD$y$ calculated by SRW. (d) The intensity calculated by coherent modes. (e) The CSD$x$ calculated by coherent modes. (f) The CSD$y$ calculated by coherent modes. (g) The comparison of 1d intensity (summed along horizontal and vertical direction) calculated by SRW and coherent modes. (h) The comparison of CSD$x$ calculated by SRW and coherent modes. (i) The comparison of CSD$y$ calculated by SRW and coherent modes. SRW represents the results calculated by SRW

software. CM represents the results calculated by decomposed coherent modes.

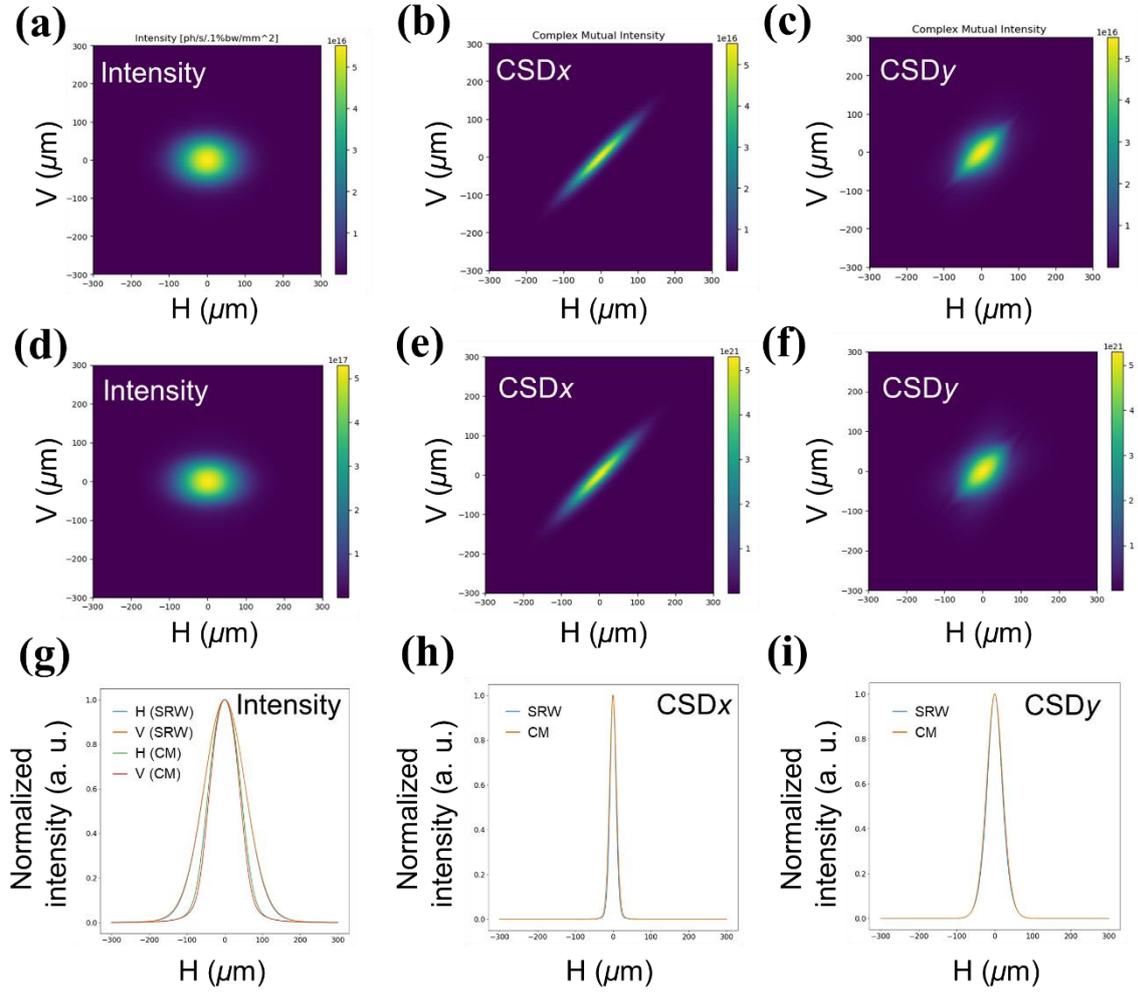

**Figure. S2.** The comparison based on NAMI beamline (12.4keV, **Table 1**, **column NAMI**). (a) The intensity calculated by SRW. (b) The CSD$x$ calculated by SRW. (c) The CSD$y$ calculated by SRW. (d) The intensity calculated by coherent modes. (e) The CSD$x$ calculated by coherent modes. (f) The CSD$y$ calculated by coherent modes. (g) The comparison of 1d intensity (summed along horizontal and vertical direction) calculated by SRW and coherent modes. (h) The comparison of CSD$x$ calculated by SRW and coherent modes. (i) The comparison of CSD$y$ calculated by SRW and coherent modes. "SRW" means the results calculated by SRW software. "CM" means the results calculated by decomposed coherent modes.

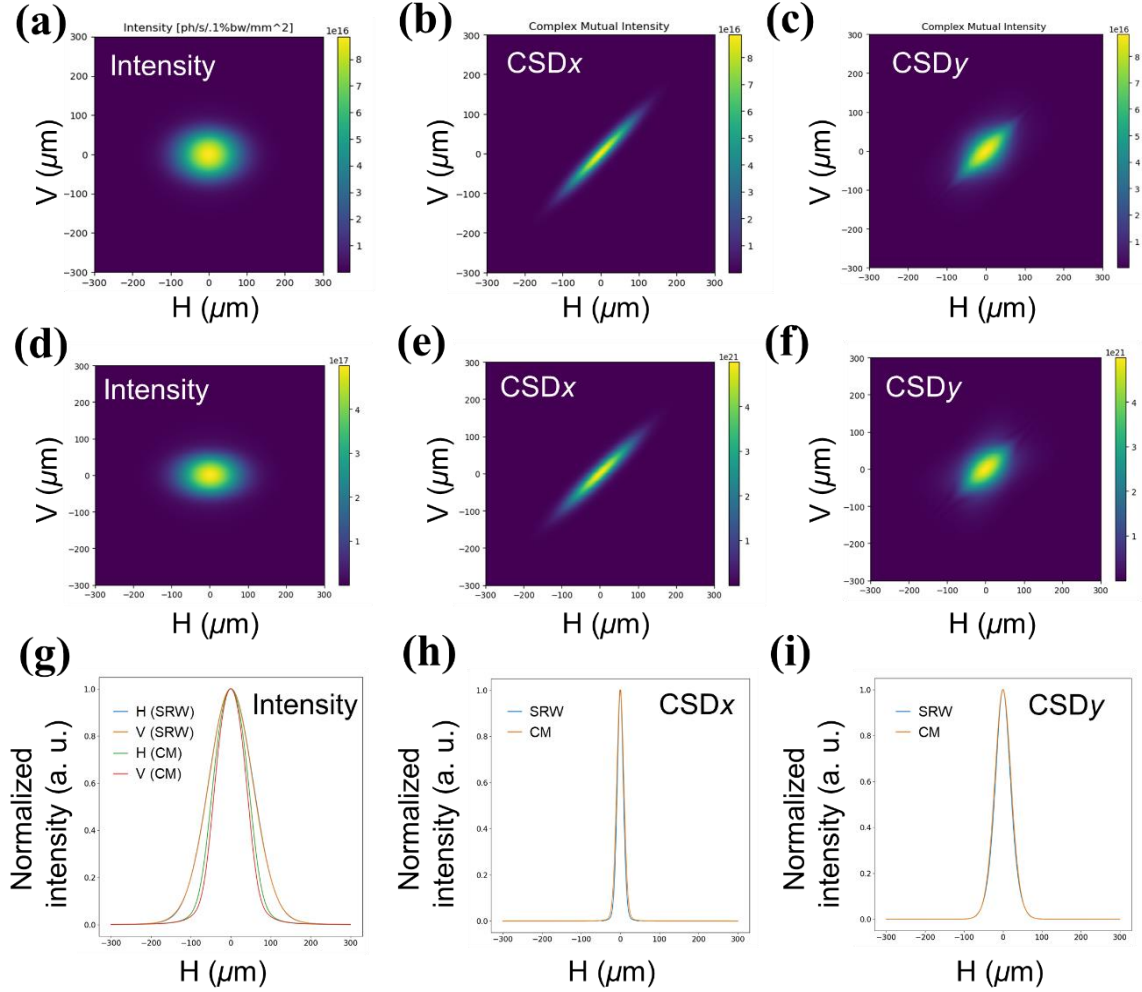

**Figure. S3.** The comparison based on HXCS beamline (12.4keV, Table 1, column HXCS). (a) The intensity calculated by SRW. (b) The CSD*x* calculated by SRW. (c) The CSD*y* calculated by SRW. (d) The intensity calculated by coherent modes. (e) The CSD*x* calculated by coherent modes. (f) The CSD*y* calculated by coherent modes. (g) The comparison of 1d intensity (summed along horizontal and vertical direction) calculated by SRW and coherent modes. (h) The comparison of CSD*x* calculated by SRW and coherent modes. (i) The comparison of CSD*y* calculated by SRW and coherent modes. "SRW" means the results calculated by SRW software. "CM" means the results calculated by decomposed coherent modes.

The coherent mode decomposition of ESRF-EBS has been performed by *Comsyl* (Sanchez del Rio *et al.*, 2019). Here, the decomposed coherent modes and the coherent fraction of ESRF-EBS were also calculated by the methods used in this article. The calculated coherent fraction was 2.94%, which was consistent with the reported result (2.8%). The decomposed coherent modes were propagated to source by 1:1 focusing. The results are similar with the reported results (**Figure 9** of Sanchez del Rio *et al.*, 2019).

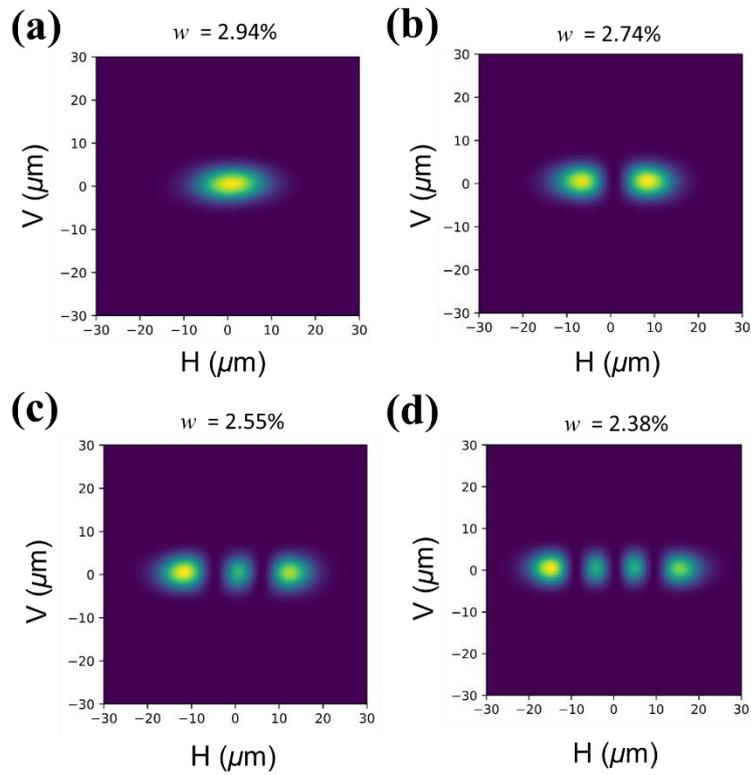

**Figure. S4.** The decomposed coherent modes of ESRF-EBS (m = 0~3).

## S2. Verification of propagators.

The propagators of the model were verified by focusing and diffracting a wavefront intensity. The results calculated by SRW and the propagators used in this article were compared. Firstly, a wavefront was calculated by SRW (Table 1, HXCS column) from source to an optical plane (distance is 10 m). Then, this wavefront was focused by an ideal lens (focus length = 5 m, 1: 1 focus). For diffraction, the size of the slit was 50 $\mu m$，and the distance between the slit and final screen is 10 m.

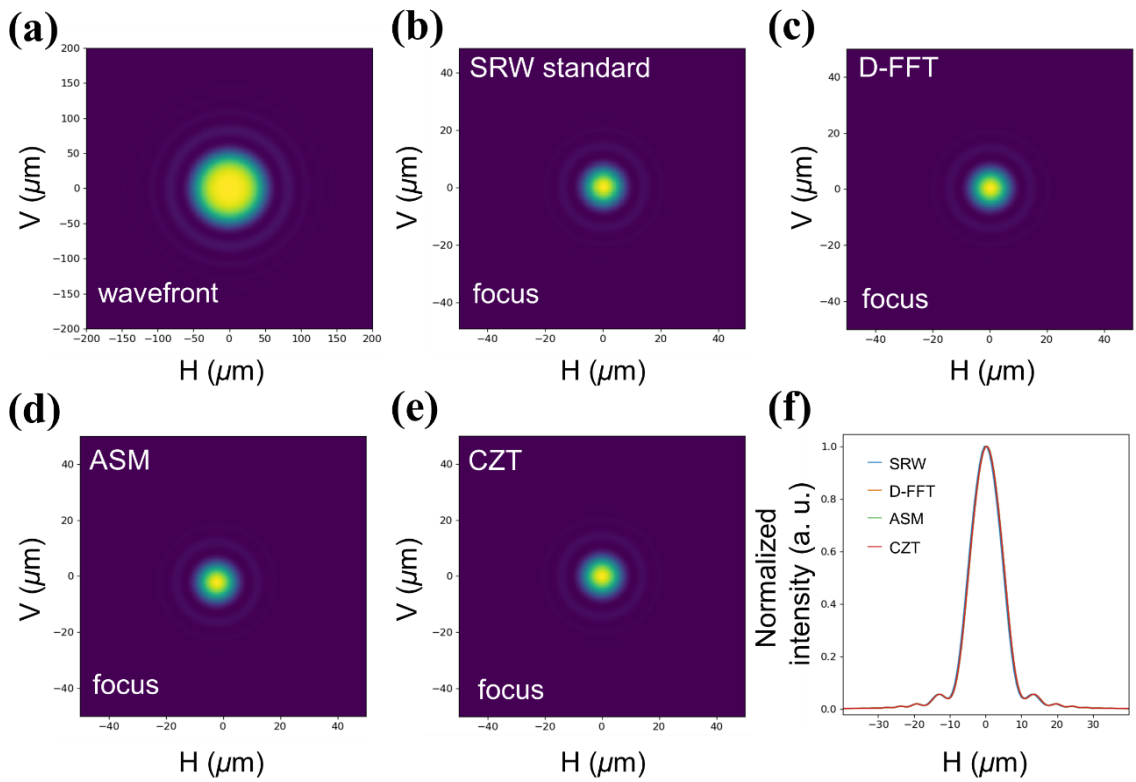

**Figure. S5.** (a) The wavefront calculated by SRW. (b) The focus spots calculated by SRW standard propagator, (c) Fresnel diffraction integral (D-FFT algorithm), (d) angular spectrum method (ASM), and (e) chirp z transform method (CZT) were compared. (f) The 1d results (summed along vertical direction) were shown.

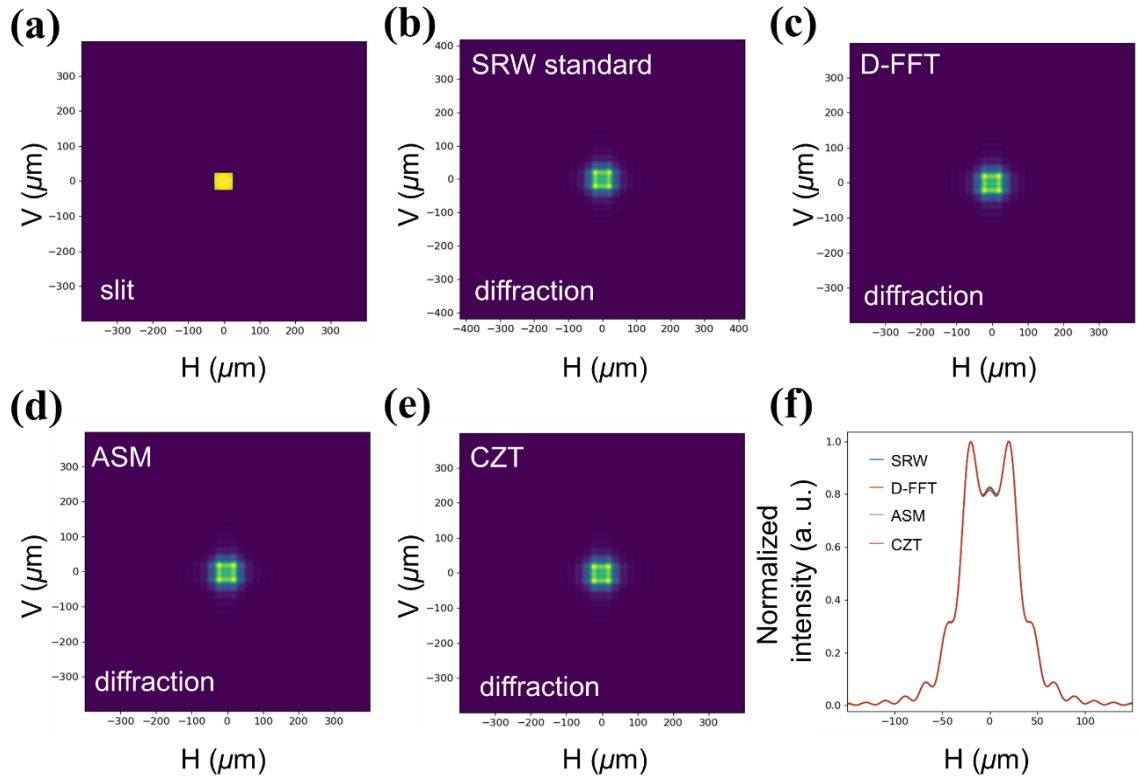

**Figure. S6.** (a) The wavefront after slit that calculated by SRW. (b) The diffraction patterns calculated by SRW standard propagator, (c) Fresnel diffraction integral (D-*FFT* algorithm), (d) angular spectrum method (ASM), and (e) chirp z transform method (CZT) were compared. (f) The 1d results (summed along vertical direction) were shown.

**S3. Propagation of coherent modes based on the optical layout of 100 nm focusing.**

The 100 nm focusing optical layout of HXCS is shown in Figure. S7a. Advanced-KB mirror (AKB, 1D Wolter mirror) was used to focus the X-ray beam to 85 μm×97 μm. The coherence slit was located before AKB mirror, and the size of the slit is 159.7 μm×398.1 μm (Horizontal×Vertical). Here, the propagation of the first coherent mode was shown as an example, The D-*FFT* propagator was used to propagate the coherent mode from source to coherence slit. The propagation of coherent mode within the AKB mirror (two pairs of elliptical and hyperbolic mirrors) used an ASM propagator. Then, the focusing from AKB mirror to focus spot used CZT propagator.

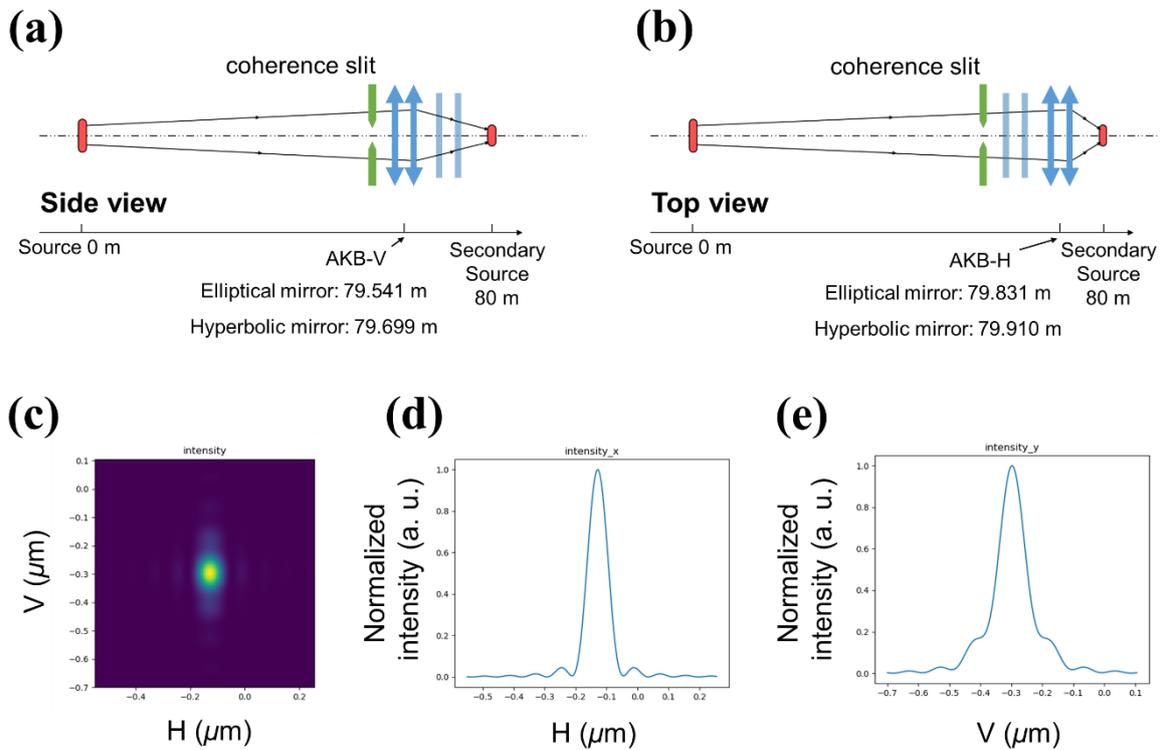

**Figure. S7.** (a) The side and (b) top views of the 100 nm focusing optical layout of HXCS were shown. (c) The two-dimensional focus spot. The summation of focus spots along along (d) vertical and (e) horizontal direction.